\documentclass[aps,prd,onecolumn,groupedaddress,showpacs,nofootinbib,amssymb]{revtex4}
\usepackage[dvips]{graphicx}
\usepackage{amssymb}
\usepackage{amsmath}
\usepackage{graphicx,,color}
\usepackage{amsfonts}
\usepackage{bm}
\usepackage{cancel}
\usepackage{comment}
\usepackage{listings}

\newcommand\be{\begin{equation}}
\newcommand\ee{\end{equation}}

\allowdisplaybreaks[4]

\begin{document}

\title{Effective Equation of State Oscillations at Matter-Radiation Equality and Primordial Gravitational Waves}
\author{S.D. Odintsov$^{1,2}$}\email{odintsov@ice.csic.es}
\author{V.K. Oikonomou,$^{3,4}$}\email{voikonomou@gapps.auth.gr;v.k.oikonomou1979@gmail.com}
\affiliation{$^{1)}$ Institute of Space Sciences (ICE, CSIC) C. Can Magrans s/n, 08193 Barcelona, Spain \\
$^{2)}$ ICREA, Passeig Luis Companys, 23, 08010 Barcelona, Spain\\
$^{3)}$Department of Physics, Aristotle University of
Thessaloniki, Thessaloniki 54124, Greece\\
$^{4)}$L.N. Gumilyov Eurasian National University - Astana,
010008, Kazakhstan}

\tolerance=5000

\begin{abstract}
The theory controlling the Universe's evolution in the classical
regime has to be motivated by particle physics reasoning and
should also generate inflation and dark energy eras in a unified
way. One such framework is $F(R)$ gravity. In this work we examine
a class of exponential deformations of $R^2$ gravity motivated by
fundamental physics of scalaron evolution in a de Sitter
background. As we show this class of models describe both
inflation and the dark energy era in a viable way compatible with
the Planck constraints on inflation and the cosmological
parameters. Regarding the inflationary era, the exponentially
deformed $R^2$ model also yields a rescaled Einstein-Hilbert term
which remarkably does not affect the dynamics and the inflationary
evolution is identical to that of an $R^2$ model. The dark energy
era is also found to be viable and mimics the
$\Lambda$-Cold-Dark-Matter model. More importantly, this class of
$F(R)$ gravity exponential $R^2$ deformations also has an
important characteristic, and specifically it yields total
equation of state oscillations deeply in the matter domination
era, for redshifts $z\sim 3400$, so near the matter-radiation
equality. These total equation of state deformations at such a
large redshift may directly affect the energy spectrum of the
primordial gravitational waves. Indeed as we show, the effect is
measurable and it leads to an enhancement of the tensor
perturbations energy spectrum for low frequencies probed by the
future LiteBIRD mission. This enhancement might have a measurable
effect on the $B$-modes of the Cosmic Microwave Background
radiation and thus may be detectable by the LiteBIRD mission. Only
a handful of theoretical frameworks can generate the gravitational
wave pattern generated by the class of exponentially deformed
$R^2$ models we presented.
\end{abstract}


\maketitle

\section{Introduction}

The mysteries that remain unanswered in modern theoretical physics
are the dark sector and the primordial era of the Universe, where
inflation is the theorized description for this era. The dark
sector is composed by dark matter
\cite{Bertone:2004pz,Bergstrom:2000pn,Mambrini:2015sia,Profumo:2013yn,Hooper:2007qk,Oikonomou:2006mh,Caputo:2024oqc,Kuster:2008zz,Archidiacono:2013cha}
and dark energy. Dark energy is the observed accelerating
evolution of the Universe, firstly confirmed in the late 90's
\cite{Riess:1998cb} and is currently studied in many theoretical
physics contexts. The standard description of dark energy consists
of the cosmological constant description, and specifically the
$\Lambda$-Cold-Dark-Matter ($\Lambda$CDM) model has several
successes to date. However, there are shortcomings of the
$\Lambda$CDM, such as the possibility that dark energy is
dynamical, as indicated by the 2024 DESI data \cite{DESI:2024mwx}.
Also the Planck data allow for phantom dark energy, which cannot
be described in standard general relativity without using phantom
scalar fields. Apart from the above mentioned issues of the
$\Lambda$CDM description of dark energy, an appealing theoretical
description of our Universe should describe inflation and the
late-time acceleration era in a unified way. The only framework
that can consistently achieve this is $F(R)$ gravity, see the
pioneer article \cite{Nojiri:2003ft} and also Refs.
\cite{Nojiri:2007as,Nojiri:2007cq,Cognola:2007zu,Nojiri:2006gh,Appleby:2007vb,Elizalde:2010ts,Odintsov:2020nwm,Sa:2020fvn,Appleby:2009uf}
and the reviews
\cite{reviews1,reviews2,reviews3,reviews4,reviews5}. The $F(R)$
gravity theoretical framework offers the unique possibility of
describing simultaneously inflation and dark energy and also the
dark energy era of $F(R)$ gravity is dynamical, while evolutionary
the $F(R)$ gravity dark energy era mimics the $\Lambda$CDM model
at late times.

However, it is known that in the context of $F(R)$ gravity
dynamical dark energy, the total equation of state (EoS)
experiences strong oscillations. In this work we develop a class
of $F(R)$ gravity models which can generate a viable inflationary
era and in addition a viable dark energy era, with the same model.
We numerically solve the field equations and prove that the total
EoS experiences strong oscillations up to redshifts $z\sim 3400$
so in the beginning of the dark matter era. The total EoS
parameter oscillates in the range $w_{tot}\sim [0.13-0.2]$ and
this total EoS oscillation can have observable imprints on the
energy spectrum of the primordial gravitational waves. Indeed, as
we demonstrate the total EoS oscillation affects the energy
spectrum and it generates an enhancement of the tensor spectrum
for frequencies that will be probed by the LiteBIRD
\cite{LiteBIRD:2022cnt} mission. This enhancement of the tensor
spectrum for such low frequencies can affect directly the Cosmic
Microwave Background (CMB) radiation and therefore it can be
detected in the B-mode spectrum of the CMB, which can be detected
by the LiteBIRD mission.

This article is organized as follows: In section II we present a
class of $F(R)$ gravity models which can generate a viable
inflationary era primordially, while simultaneously it can
generate a viable dark energy era. For the dark energy era study
we solve numerically the field equations using appropriate initial
conditions that go deeply in the dark matter domination era,
nearly at the beginning of the matter domination era. As we show,
the unification $F(R)$ gravity model can generate a successful
dark energy era compatible with the latest Planck constraints
\cite{Aghanim:2018eyx} on the cosmological parameters. We will
explicitly show the oscillations of the total EoS parameter deeply
in the matter domination era and in section II we shall analyze
the effects of the total EoS oscillations on the energy spectrum
of the primordial waves. Also we shall discuss why the resulting
enhancement of the tensor spectrum for frequencies probed by the
LiteBIRD can enhance the B-modes of the CMB and can therefore be
detected by LiteBIRD in 2029 and beyond.

For the purposes of this article we shall assume that the
spacetime is described by a flat Friedmann-Robertson-Walker (FRW)
metric of the form,
\begin{equation}
\label{metricfrw} ds^2 = - dt^2 + a(t)^2 \sum_{i=1,2,3}
\left(dx^i\right)^2\, ,
\end{equation}
where $a(t)$ is the scale factor, and the Hubble rate is
$H=\frac{\dot{a}}{a}$, while the Ricci scalar is
$R=12H^2+6\dot{H}$.

\section{An Class of $F(R)$ Gravity Models Providing a Unified Description of Dark Energy and Inflation}

In this section we shall present a class of exponential
deformations of the $R^2$ model that may lead to a unified
description of inflation and the dark energy era. These models
have a deeper origin and are not some models chosen
phenomenologically. As we show in \cite{mywork}, these models stem
from a deeper theoretical reasoning having to do with the scalaron
mass in a de Sitter background. Apart from the unification of the
inflationary era with the dark energy era, the exponential
deformation of the $R^2$ model we shall consider also generates
strong total EoS oscillations near the epoch of matter-radiation
equality, and as it proves, this can have measurable effects on
the energy spectrum of the primordial gravitational waves.
Consider the following action, which describes $F(R)$ gravity in
the presence of matter perfect fluids,
\begin{equation} \label{mainaction}
\mathcal{S}=\int d^4x\sqrt{-g}\left(
\frac{1}{2\kappa^2}F(R)+\mathcal{L}_m \right)\, ,
\end{equation}
where $\kappa^2=\frac{1}{8\pi G}=\frac{1}{M_p^2}$, and $G$ is
Newton's gravitational constant while $M_p$ stands for the reduced
Planck mass. The Lagrangian density $\mathcal{L}_m$ denotes the
perfect matter fluids presents, which we will assume that are
composed by radiation and dust. The class of exponential
deformations of the $R^2$ model has the following form,
\begin{equation}\label{effectivelagrangian1initial}
F(R)=R+\frac{R^2}{M^2}+\frac{R^2}{M^2}\,e^{\frac{\gamma\Lambda}{R}}+\lambda\,R\,e^{\frac{\gamma\Lambda}{R}}-50\lambda
\Lambda -\frac{\Lambda}{\zeta}
\left(\frac{R}{m_s^2}\right)^{\delta }\, ,
\end{equation}
where $\lambda$, $\zeta$ and $\gamma$ are dimensionless
parameters. These models provide a unified description of
inflation and the dark energy era as we evince in this section.
Before we get into the details of the inflationary era for these
models, we shall discuss in a nutshell why the phenomenological
class of exponential deformations of the $R^2$ model appearing in
Eq. (\ref{effectivelagrangian1initial}) are physically motivated
from first principles, and they also enable the unification of the
inflationary era with the dark energy era. The full details of
this analysis will appear in Ref. \cite{mywork} and here we give a
brief account. The de Sitter solution existence criterion can be
found by perturbing the field equations of $F(R)$ gravity for a
FRW spacetime. Specifically, consider $R=R_0+G(R)$, with $R_0$
being the scalar curvature of the de Sitter solution, then, the
scalaron field $\mathcal{G}=F'(R)$ in the Einstein frame satisfies
the equation,
\begin{equation}\label{scalaronequation}
\square \mathcal{G}+m^2 \mathcal{G}=0\, ,
\end{equation}
where the scalaron mass is \cite{Muller:1987hp},
\begin{equation}\label{scalaronmassinitial}
m^2=\frac{1}{3}\left(-R+\frac{F_R}{F_{RR}} \right)\, ,
\end{equation}
Now, as we show in \cite{mywork}, we require two physical
condition, firstly that $m^2\geq 0$ for all curvatures, and also
that the scalaron mass decreases as the curvature decreases, in a
monotonic way. The reason for the latter requirement is that at
late times we basically require that the scalaron mass is smaller
than the early times scalaron mass. This will enable the
unification of the dark energy with the inflationary era in a
natural way. In order to have monotonically decreasing scalaron
mass, one must require that,
\begin{equation}\label{derivativescalaronmass}
\frac{\partial m^2(R)}{\partial R}\geq 0 \, ,
\end{equation}
so effectively we must have,
\begin{equation}\label{equationderivativescalaron}
\frac{\partial m^2(R)}{\partial
R}=-\frac{1}{12}\frac{F_R}{R\,F_{RR}}\,\frac{4\,R\,F_{RRR}}{F_{RR}}\geq
0 \, .
\end{equation}
The parameter $x$ defined as,
\begin{equation}\label{parameterx}
x=\frac{4 F_{RRR}\,R}{F_{RR}}\, ,
\end{equation}
plays a fundamental role in this analysis and it proves that for
the class of models of Eq. (\ref{effectivelagrangian1initial}) the
parameter $x$ is approximately zero and negative, which ensures a
viable inflationary era. At the same time, due to the fact that
the de Sitter scalaron mass satisfies $\frac{\partial
m^2(R)}{\partial R}\geq 0$, the unification of the dark energy
with the inflationary era is achieved. This is a unique
characteristic of the class of models of Eq.
(\ref{effectivelagrangian1initial}) as we show in \cite{mywork}.

Let us first start with the inflationary era. During the
primordial inflation era, the Hubble rate has values of the order
$H_I=10^{13}$GeV hence the Ricci scalar $R\sim H_I^2$ takes quite
large values, thus primordially the $F(R)$ gravity which drives
the evolution is at leading order,
\begin{equation}\label{effectivelagrangian1}
F(R)\simeq (1+\lambda)
R+\frac{R^2}{M^2}+\mathcal{O}(\frac{\Lambda}{R})\, ,
\end{equation}
hence this is a rescaled version of Einstein-Hilbert gravity in
the presence of an $R^2$ term, with the rescaling term being
quantified by the parameter $\lambda$. However, as we now show in
brief, the rescaling does not affect the inflationary era and the
evolution is described by a quasi-de Sitter evolution generated by
the $R^2$ term. Let us show this in brief, so for the $F(R)$
gravity of Eq. (\ref{effectivelagrangian1}), the Friedmann
equation becomes,
\begin{equation}\label{diffeqndefomedstaro}
\ddot{H}+3 H \dot{H}-\frac{\dot{H}^2}{2 H}+\frac{1}{12} \lambda
M^2 H+\frac{1}{12} M^2 H=0\, .
\end{equation}
Using the slow-roll approximation for the inflationary era,
\begin{equation}\label{slowrollapproximation}
\ddot{H}\ll H \dot{H},\,\,\,\dot{H}\ll H^2\, ,
\end{equation}
the Friedmann equation (\ref{diffeqndefomedstaro}) takes the form,
\begin{equation}\label{finalformfriedmanneqn}
\dot{H}\simeq -\frac{1}{36} (\lambda +1) M^2\, ,
\end{equation}
and its solution is a quasi-de Sitter evolution of the form,
\begin{equation}\label{quasidesitter}
H(t)=H_I-\frac{1}{36} t \left(\lambda  M^2+M^2\right)\, ,
\end{equation}
with $H_I$ being an integration constant, the inflationary scale.
The inflationary phenomenology can be found if one calculates the
slow-roll parameters, defined as follows
\cite{Hwang:2005hb,reviews1,Odintsov:2020thl},
\begin{equation}
\label{restofparametersfr}\epsilon_1=-\frac{\dot{H}}{H^2}, \quad
\epsilon_2=0\, ,\quad \epsilon_3= \frac{\dot{F}_R}{2HF_R}\, ,\quad
\epsilon_4=\frac{\ddot{F}_R}{H\dot{F}_R}\,
 .
\end{equation}
The observational indices of inflation are defined in terms of the
slow-roll parameters as follows \cite{Odintsov:2020thl},
\begin{equation}
\label{spectralfinal} n_s\simeq 1-6\epsilon_1-2\epsilon_4\, ,
\end{equation}
\begin{equation}
\label{tensorfinal} r\simeq 48\epsilon_1^2\, .
\end{equation}
The parameter $\epsilon_4$ is \cite{Odintsov:2020thl},
\begin{equation}\label{finalapproxepsilon4}
\epsilon_4\simeq -\frac{24
F_{RRR}H^2}{F_{RR}}\epsilon_1-\epsilon_1\, ,
\end{equation}
therefore the spectral index of the primordial scalar
perturbations and the tensor-to-scalar ratio is,
\begin{equation}
\label{spectralfinal} n_s\simeq 1-4\epsilon_1\, ,
\end{equation}
\begin{equation}
\label{tensorfinal} r\simeq 48\epsilon_1^2\, .
\end{equation}
Hence, the calculation of the first slow-roll index is central in
our analysis. This can be evaluated easily for the quasi-de Sitter
evolution at hand (\ref{finalformfriedmanneqn}), and it is equal
to,
\begin{equation}\label{epsilon1indexanalytic}
\epsilon_1=-\frac{-\lambda  M^2-M^2}{36 \left(H_I-\frac{1}{36} t
\left(\lambda M^2+M^2\right)\right)^2}\, .
\end{equation}
Having this at hand, we can evaluate the time instances at which
inflation starts and ends. By solving the equation
$\epsilon_1(t_f)=1$ we can find the time instance $t_f$ at which
inflation ends, which is,
\begin{equation}\label{finaltimeinstance}
t_f=\frac{6 \left(6 H_I \lambda  M^2+6 H_I M^2-\sqrt{\lambda ^3
M^6+3 \lambda ^2 M^6+3 \lambda M^6+M^6}\right)}{\lambda ^2 M^4+2
\lambda  M^4+M^4}\, .
\end{equation}
Now using the definition of the $e$-foldings number $N$,
\begin{equation}\label{efoldingsnumber}
N=\int_{t_i}^{t_f}H(t)dt\, ,
\end{equation}
we can find the time instance  $t_i$ at which inflation starts. So
by replacing the quasi de-Sitter evolution in the $e$-foldings
number, we get,
\begin{equation}\label{ti}
t_i=\frac{6 \left(6 H_I+\sqrt{(\lambda +1) M^2 (2
N+1)}\right)}{(\lambda +1) M^2}\, ,
\end{equation}
Now we can evaluate the first slow-roll index $\epsilon_1$ at
first horizon crossing, and we have,
\begin{equation}\label{epsilon1lambdaind}
\epsilon_1(t_i)=\frac{1}{1+2N}\, ,
\end{equation}
hence expanding the spectral index and the tensor-to-scalar ratio
at leading order in the parameter $N$, we obtain $n_s\sim
1-\frac{2}{N}$ and $r\sim \frac{12}{N^2}$, which note that are
identical to the observational indices of the pure $R^2$ model
with $\lambda=0$, which is well compatible with the Planck 2018
data \cite{Aghanim:2018eyx}. Remarkably, the rescaling of the
Einstein-Hilbert term does not affect at all the inflationary
phenomenology of the model. Having discussed the inflationary era,
now let us turn our focus to the dark energy era. This is done in
the next subsection.

\subsection{Total EoS Oscillations at Matter-Radiation Equality}

\begin{figure}
\centering
\includegraphics[width=18pc]{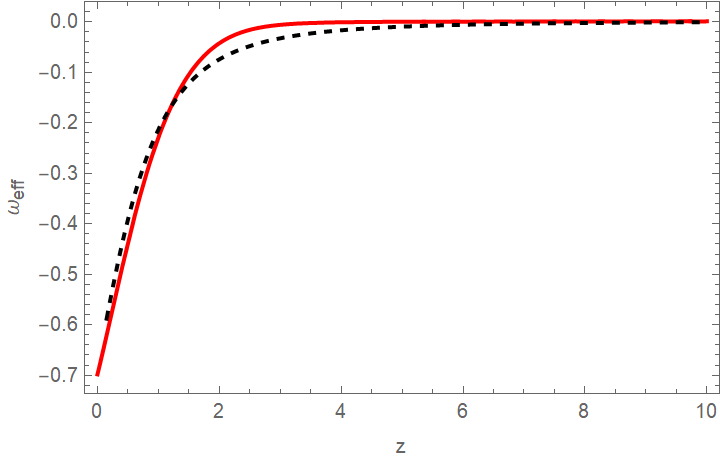}
\includegraphics[width=18pc]{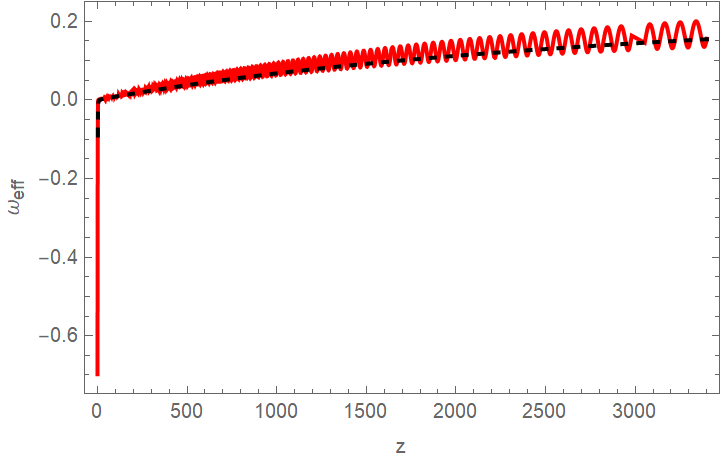}
\includegraphics[width=18pc]{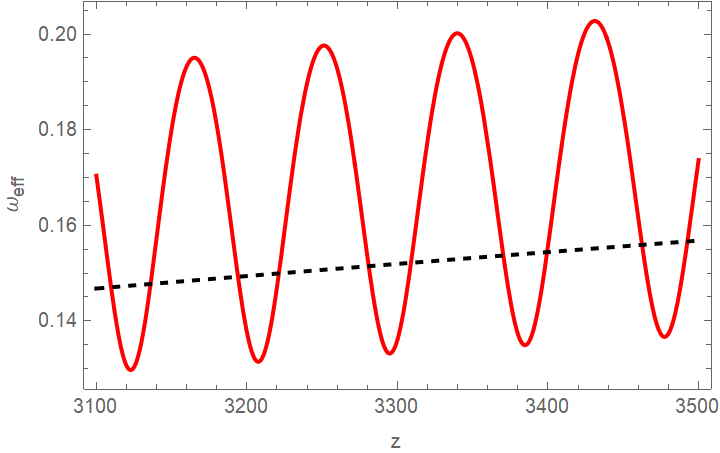}
\caption{The total EoS parameter $\omega_{eff}$ as a function of
the redshift for the $\Lambda$CDM model (black dashed curves) and
the $F(R)$ gravity model (red curves).}\label{plot2}
\end{figure}

In this subsection we shall consider the evolution of the Universe
driven by the $F(R)$ gravity of Eq.
(\ref{effectivelagrangian1initial}). We shall be interested in the
evolution from the beginning of the matter domination era, until
the present day dark energy era, and we shall confront the model
with the latest Planck constraints on the cosmological parameters.
Interestingly enough, the $F(R)$ model
(\ref{effectivelagrangian1initial}) leads to strong total EoS
oscillations at a redshift $z\sim 3400$ which in turn can have
observable imprints on the energy spectrum of the primordial
gravitational waves. To start with, the field equations of $F(R)$
gravity in a flat FRW background can be written in the form of
Einstein-Hilbert gravity, in the following way,
\begin{align}\label{flat}
& 3H^2=\kappa^2\rho_{tot}\, ,\\ \notag &
-2\dot{H}=\kappa^2(\rho_{tot}+P_{tot})\, ,
\end{align}
with $\rho_{tot}=\rho_{m}+\rho_{G}+\rho_r$ standing for the total
energy density, $\rho_m$ denotes the cold dark matter energy
density and $\rho_r$ stands for the radiation energy density.
Finally, $\rho_{G}$ stands for the energy density of the geometric
fluid, which is due to the $F(R)$ gravity,
\begin{equation}\label{degeometricfluid}
\kappa^2\rho_{G}=\frac{F_R R-F}{2}+3H^2(1-F_R)-3H\dot{F}_R\, .
\end{equation}
In addition, $P_{tot}=P_r+P_{G}$ stands for the total pressure of
the cosmological fluid, and the geometric pressure is,
\begin{equation}\label{pressuregeometry}
\kappa^2P_{G}=\ddot{F}_R-H\dot{F}_R+2\dot{H}(F_R-1)-\kappa^2\rho_{G}\,
.
\end{equation}
We shall solve numerically the field equations using appropriate
initial conditions that will go deeply in the matter domination
era. Also we shall study the behavior of several statefinder
parameters and we shall examine the behavior of the total EoS
parameter to reveal the oscillations at the early stages of the
matter domination era. We shall express the field equations in
terms of the redshift $z$ defined as,
\begin{equation}\label{redshift}
1+z=\frac{1}{a}\, ,
\end{equation}
where we assumed that the present time scale factor is equal to
unity. Also we shall express the field equations in terms of the
statefinder function $y_H(z)$
\cite{Hu:2007nk,Bamba:2012qi,reviews1} defined as,
\begin{equation}\label{yHdefinition}
y_H(z)=\frac{\rho_{G}}{\rho^{(0)}_m}\, ,
\end{equation}
where $\rho^{(0)}_m$ stands for the energy density of cold dark
matter at present time. Then, the statefinder $y_H(z)$ is written
as,
\begin{equation}\label{finalexpressionyHz}
y_H(z)=\frac{H^2}{m_s^2}-(1+z)^{3}-\chi (1+z)^4\, ,
\end{equation}
where $\chi=\frac{\rho^{(0)}_r}{\rho^{(0)}_m}\simeq 3.1\times
10^{-4}$, and also $\rho^{(0)}_r$ is the radiation energy density,
and in addition
$m_s^2=\frac{\kappa^2\rho^{(0)}_m}{3}=H_0\Omega_c=1.37201\times
10^{-67}$eV$^2$. The Friedmann equation in terms of the
statefinder $y_H(z)$ becomes \cite{Bamba:2012qi},
\begin{equation}\label{differentialequationmain}
\frac{d^2y_H(z)}{d z^2}+J_1\frac{d y_H(z)}{d z}+J_2y_H(z)+J_3=0\,
,
\end{equation}
and we introduced the dimensionless functions $J_1$, $J_2$ and
$J_3$ which are,
\begin{align}\label{diffequation}
& J_1=\frac{1}{z+1}\left(
-3-\frac{1-F_R}{\left(y_H(z)+(z+1)^3+\chi (1+z)^4\right) 6
m_s^2F_{RR}} \right)\, , \\ \notag & J_2=\frac{1}{(z+1)^2}\left(
\frac{2-F_R}{\left(y_H(z)+(z+1)^3+\chi (1+z)^4\right) 3
m_s^2F_{RR}} \right)\, ,\\ \notag & J_3=-3(z+1)-\frac{\left(1-F_R
\right)\Big{(}(z+1)^3+2\chi (1+z)^4
\Big{)}+\frac{R-F}{3m_s^2}}{(1+z)^2\Big{(}y_H(z)+(1+z)^3+\chi(1+z)^4\Big{)}6m_s^2F_{RR}}\,
,
\end{align}
with $F_{RR}=\frac{\partial^2 F}{\partial R^2}$. Our strategy is
to solve numerically the Friedmann equation
(\ref{differentialequationmain}) using initial conditions that
will go deeply in the matter domination era, up to present day.
Regarding the initial conditions, here is our way of thinking and
how we will choose the initial conditions.
\begin{figure}
\centering
\includegraphics[width=18pc]{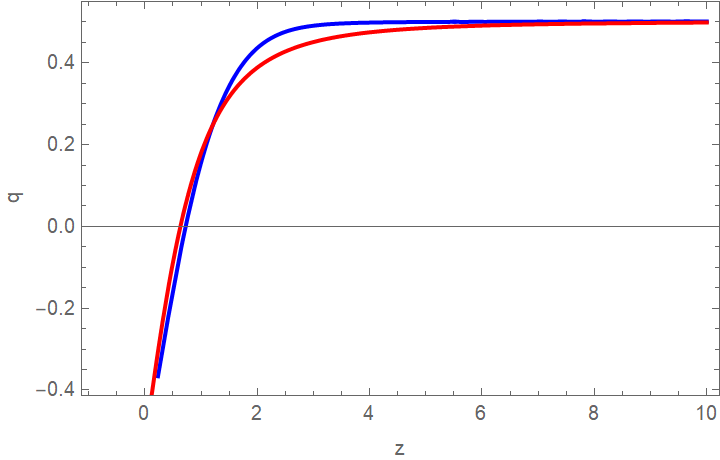}
\includegraphics[width=18pc]{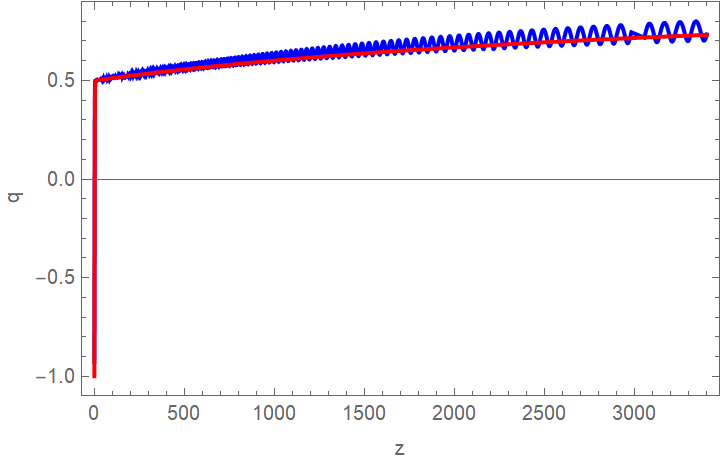}
\caption{The deceleration parameter, as a function of the redshift
for the $\Lambda$CDM model (red curves) and the $F(R)$ gravity
model (blue curves) for various redshift ranges.}\label{plot1}
\end{figure}
Deeply in the matter domination era, the curvature would be
$R=3m_s^2(1+z)^3$, thus we choose the initial redshift to be
$z_f=3500$ and we choose the following initial conditions for the
statefinder $y_H(z)$:
\begin{equation}\label{generalinitialconditions}
y_H(z_f)=\frac{\Lambda}{3m_s^2}\left(
1-\frac{(1+z_f)^3}{1000}\right)\, , \,\,\,\frac{d y_H(z)}{d
z}\Big{|}_{z=z_f}=-\frac{1}{1000}\frac{\Lambda}{3m_s^2}(1+z_f)^2\,
.
\end{equation}
We have developed a numerical code which is appropriately
constructed to integrate backwards the differential equation
(\ref{differentialequationmain}) from $z=3500$ to present day
$z=0$. This is a PYTHON 3 based code, and specifically it utilizes
the $``\mathrm{solve}_{\_ }\mathrm{ivp}''$ function of the SCIPY
module and also it is an LSODA solver. The code can be found here
\cite{code}. For our numerical analysis, it is vital to express
several quantities in terms of the statefinder $y_H(z)$ because we
will need these. Firstly the curvature is expressed as follows,
\begin{equation}\label{ricciscalarasfunctionofz}
R(z)=3m_s^2\left( 4y_H(z)-(z+1)\frac{d y_H(z)}{d
z}+(z+1)^3\right)\, ,
\end{equation}
and also the dark energy density parameter $\Omega_{DE}$ is
expressed as,
\begin{equation}\label{omegaglarge}
\Omega_{DE}(z)=\frac{y_H(z)}{y_H(z)+(z+1)^3+\chi (z+1)^4}\, .
\end{equation}
while the dark energy EoS parameter is,
\begin{equation}\label{omegade}
\omega_{DE}(z)=-1+\frac{1}{3}(z+1)\frac{1}{y_H(z)}\frac{d
y_H(z)}{d z}\, .
\end{equation}
Moreover, an important parameter which shall be extensively used
regarding its effects on the energy spectrum of  the primordial
gravitational waves is the total EoS parameter, which in terms of
the statefinder $y_H(z)$ reads,
\begin{equation}\label{totaleosparameter}
\omega_{tot}(z)=\frac{2 (z+1) H'(z)}{3 H(z)}-1\, .
\end{equation}
In addition, the deceleration parameter takes the form,
\begin{align}\label{statefinders}
& q=-1-\frac{\dot{H}}{H^2}=-1+(z+1)\frac{H'(z)}{H(z)}\, .
\end{align}
Also for the whole duration of the integration, we shall compare
our results of the $F(R)$ gravity evolution, with the $\Lambda$CDM
model, the Hubble rate of which is equal to,
\begin{equation}\label{lambdacdmhubblerate}
H_{\Lambda}(z)=H_0\sqrt{\Omega_{\Lambda}+\Omega_M(z+1)^3+\Omega_r(1+z)^4}\,
,
\end{equation}
where $H_0$ is the Hubble rate at present time, which is
$H_0\simeq 1.37187\times 10^{-33}$eV according to the 2018 Planck
data \cite{Aghanim:2018eyx}. Also $\Omega_{\Lambda}\simeq
0.681369$ and in addition $\Omega_M\sim 0.3153$
\cite{Aghanim:2018eyx}, and furthermore $\Omega_r/\Omega_M\simeq
\chi$. Regarding the numerical analysis, we shall use the
following values for the dimensionless  parameters $\delta$,
$\lambda$, $\zeta$ and $\gamma$ appearing in the $F(R)$ gravity of
Eq. (\ref{effectivelagrangian1initial}), namely, $\delta=0.9$,
$\lambda=0.0007$, $\zeta=5000$ and $\gamma=23$. Now let us proceed
to the results of our numerical analysis. In Fig. \ref{plot1} we
present the behavior of the $\Lambda$CDM deceleration parameter
(red curves) as a function of the redshift, versus the $F(R)$
gravity deceleration parameter (blue curves). We use various
redshift ranges, from $z=[0,10]$ (left plot) to $z=[0,3500]$
(right plot). As it can be seen in the two plots of Fig.
\ref{plot1}, the $F(R)$ model is very similar to the $\Lambda$CDM
in the range $z=[0,10]$ but the models are quite different deeply
in the matter domination era. This is a characteristic behavior of
the exponential $F(R)$ gravity model which shows significant
oscillations at the early stages of the matter domination era. The
same behavior is found in the dependence of the total EoS
$\omega_{eff}$ in terms of the redshift for the $\Lambda$CDM and
the $F(R)$ gravity model. This study is presented in Fig.
\ref{plot2} where we plot the total EoS parameter $\omega_{eff}$
as a function of the redshift for the $\Lambda$CDM model (black
dashed curves) and the $F(R)$ gravity model (red curves). In the
upper left plot the redshift range is $z=[0,10]$ and the behavior
of the total EoS parameter $\omega_{eff}$ for the $F(R)$ model is
almost indistinguishable from the $\Lambda$CDM model. However,
differences between the two models can be found deeply in the
matter domination era, and specifically near $z=3400$, where the
oscillations in the $F(R)$ model's total EoS parameter are
obvious, see right and bottom plots in Fig. \ref{plot2}. This can
also be seen in more details in the bottom in Fig. \ref{plot2}
where the oscillations in the total EoS of the $F(R)$ model are
more transparent. Specifically, the total EoS parameter
$\omega_{eff}$ oscillates between the values $\omega_{eff}=0.2$
and $\omega_{eff}=0.13$. We shall use this feature in the next
section where we consider the effects of these oscillations on the
energy spectrum of the primordial gravitational waves.
Before closing, let us confront the $F(R)$ gravity model's dark
energy phenomenology with the latest Planck constraints on the
cosmological parameters \cite{Aghanim:2018eyx}. Let us consider
the present day values of the dark energy density parameter
$\Omega_{DE}$ and of the dark energy EoS parameter $\omega_{DE}$
for the $F(R)$ model. Regarding $\omega_{DE}(0)$ the $F(R)$ model
yields $\omega_{DE}(0)\simeq -1.019$ so the evolution is slightly
phantom, and this is compatible with the Planck constraints which
indicate that  $\omega_{DE}=-1.018\pm 0.031$. Regarding the dark
energy density parameter for the $F(R)$ model we have
$\Omega_{DE}(0)\simeq 0.685071$ which is well fitted in the Planck
2018 constraints $\Omega_{DE}=0.6847\pm 0.0073$. Hence, the $F(R)$
model generates a viable dark energy era, and has an appealing
feature of total EoS oscillations near the beginning of the matter
domination era, which may have an observable effect on the
primordial gravitational waves as we show in the next section.

In addition, let us mention that the $F(R)$ gravity is known to
produce dark energy oscillations even from early redshifts, an
analysis firstly developed in Ref. \cite{Bamba:2012qi}. These
oscillations are due to the geometric contributions of $F(R)$
gravity to the Einstein-Hilbert action and are robust against
various initial conditions that may be used.

Also let us note at this point that $F(R)$ gravity in the
intermediate evolution epochs of our Universe, may cause some
disturbances of the EoS, as we demonstrated, but the form of the
$F(R)$ gravity we used cannot achieve a smooth transition from the
inflationary era, to the intermediate epochs before the late-time
acceleration that is, to the radiation and matter domination
epoch. The latter is generated by the cold dark matter and
baryonic matter, but still as we showed, $F(R)$ gravity may
disturb the total EoS during the matter domination epoch.

Finally, let us discuss here an interesting question, having to do
with the effects of the variation of the total EoS parameter
$\omega_{eff}$ for redshifts $z\sim 3400$, so near the
matter-radiation equality, on the CMB and the last scattering
surface. at redshift $z\sim 3400$, the Universe is at
matter-radiation equality. Thus the total EoS of the Universe is
not $\omega_{eff}=0$ but some value between the radiation EoS
$\omega_{eff}=1/3$ and the matter domination value of the total
EoS $\omega_{eff}=0$. After that it varies even until the
recombination redshift $z\sim 1100$ which is much more later. Thus
the last scattering surface will not be affected in a major way by
such variation in the total EoS for redshifts $z\sim 3400$. But
even so, these variations are taken into account in major CMB data
codes. Apart from that, in the literature an early dark energy is
also considered frequently for redshifts $z\sim 3400$, without
significantly affecting the CMB and the last scattering surface.
So no, the $\omega_{eff}=0$ condition is not correct for  $z\sim
3400$, it is not a necessity, and it is not physically correct
because that redshift corresponds to matter-radiation equality.
Let us elaborate further this issue here. During the
matter-dominated epoch in cosmology, the EoS parameter of the
universe is typically approximated as $\omega_{eff}=0$, and this
corresponds to pressureless dust, cold dark matter and baryons
contribute to the energy density. But as we mentioned,
$\omega_{eff}$ can vary slightly near the transition between
radiation domination and matter domination, i.e., around
matter-radiation equality. Specifically, before the equality,
radiation dominates $\Rightarrow \omega_{eff} \approx
\frac{1}{3}$, after the equality matter dominates, so $\Rightarrow
\omega_{eff} \approx 0$, and near the equality, both contribute,
so, $\Rightarrow \omega_{eff} \in (0, \frac{1}{3})$. This
transition is not instantaneous so the total EoS parameter
smoothly interpolates between $\frac{1}{3}$ and $0$. A question is
whether this transition affects the CMB and the last scattering
surface. The answer is yes, but slightly, and these effects have
already be taken into account in the codes related to the CMB.
Specifically, variations in the total EoS parameter affects the
sound horizon at last scattering and the angular diameter
distance, which in turn shift the CMB acoustic peaks. Also, the
decay of gravitational potentials near equality contributes to the
early ISW effect, which slightly boosts large-scale (low-$\ell$)
anisotropies in the CMB. This effect is due to the non-constant
$\omega_{eff}(a)$ around equality. As we said however, slight
deviations of the total EoS from 0 post-equality  are already
taken into account in precise CMB calculations using Boltzmann
codes like \texttt{CAMB} and \texttt{CLASS}. These codes well
incorporate the full cosmological energy budget, including
radiation, matter, and neutrinos, producing a time-dependent total
equation of state parameter $\omega_{eff}(a)$ that smoothly
transitions from $1/3$ to $0$ through equality. This evolution
affects the Hubble rate $H(a)$, the sound horizon, and thereby the
CMB acoustic peak positions \cite{Hu:2013twa}. In fact, as we
already mentioned, it is also known that even an early dark energy
era can occur near the matter-radiation equality
\cite{Chatrchyan:2024xjj}, and such an era does not have major
imprints on the last scattering surface which we need to mention
occurs much later, at $z \sim 1100$. So the variation in the EoS
around matter-radiation equality does not significantly affect the
CMB, and such effects have already been taken into account by
major observational collaborations. The $\omega_{eff}=0$ equation
of state is a theoretical approximation, not a necessity or an
accurate description of our universe during the matter-radiation
equality. Besides, as we already mentioned, at matter-radiation
equality (at redshift $z \sim 3400$), the equation of state is not
$\omega_{eff}=0$ but instead takes a value between
$\omega_{eff}=0$ and $\omega_{eff}=1/3$. The oscillatory behavior
of $\omega_{eff}(a)$ is the new feature we have highlighted and
this may affect the gravitational waves coming from inflation.

\section{Energy Spectrum Primordial Gravitational Waves and Total EoS Oscillations at $z\sim 3400$}

\begin{figure}[h!]
\centering
\includegraphics[width=40pc]{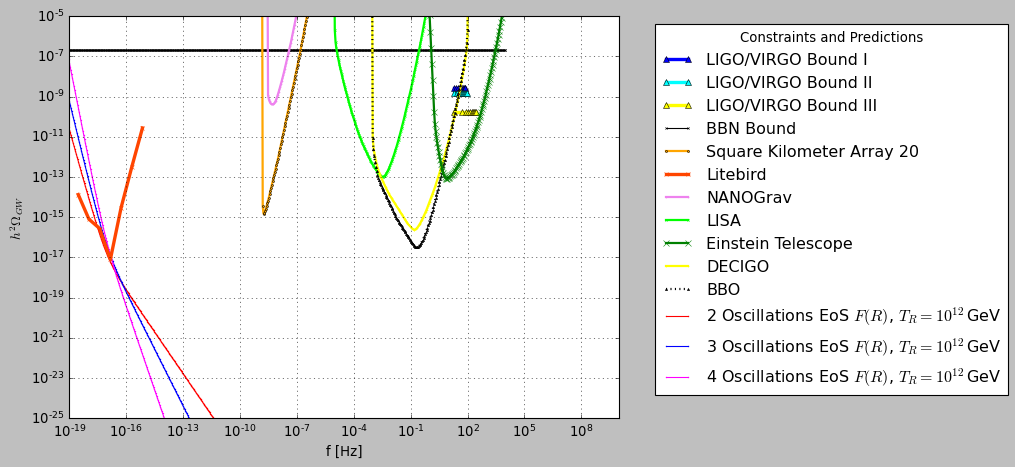}
\caption{The $h^2$-scaled gravitational wave energy spectrum for
the $F(R)$ gravity theory with $z\sim 3400$ total EoS, versus the
sensitivity curves of various current and future gravitational
waves experiments, for two, three and four oscillations of the
total EoS in the range $w=[0.13,0.20]$.} \label{FRMAIN}
\end{figure}

In this section we shall seek the imprints of the total EoS
oscillations at redshift $z\sim 3400$ on the energy spectrum of
the primordial gravitational waves. The primordial gravitational
waves issue in various contexts is heavily addressed in the
literature
\cite{Kamionkowski:2015yta,Turner:1993vb,Boyle:2005se,Zhang:2005nw,Caprini:2018mtu,Clarke:2020bil,Smith:2005mm,Giovannini:2008tm,Liu:2015psa,Vagnozzi:2020gtf,Giovannini:2023itq,Giovannini:2022eue,Giovannini:2022vha,Giovannini:2020wrx,Giovannini:2019oii,Giovannini:2019ioo,Giovannini:2014vya,Giovannini:2009kg,Kamionkowski:1993fg,Giare:2020vss,Zhao:2006mm,Lasky:2015lej,
Cai:2021uup,Odintsov:2021kup,Lin:2021vwc,Zhang:2021vak,Visinelli:2017bny,Pritchard:2004qp,Khoze:2022nyt,Casalino:2018tcd,Oikonomou:2022xoq,Casalino:2018wnc,ElBourakadi:2022anr,Sturani:2021ucg,Vagnozzi:2022qmc,Arapoglu:2022vbf,Giare:2022wxq,Oikonomou:2021kql,Gerbino:2016sgw,Breitbach:2018ddu,Pi:2019ihn,Khlopov:2023mpo,Odintsov:2022cbm,Benetti:2021uea}.
Our analysis is based on an $R^2$ inflationary era, for which the
tensor-to-scalar ratio is $r=12/N^2$ for for 60 e-foldings $r\sim
0.003$ and the tensor spectral index is $n_T=-r/8$ so it is
red-tilted. We shall consider various values of the reheating
temperature ranging from low reheating to high reheating
temperatures, however it proves that the reheating temperature
does not affect at all the phenomenology. Before we get to the
core of our analysis, it is essential to provide an account on the
energy spectrum of the primordial gravitational waves and its
detailed calculation. The energy spectrum today of the primordial
gravitational waves stemming from an inflationary theory at
present day is,
\begin{equation}
    \Omega_{\rm gw}(f)= \frac{k^2}{12H_0^2}\Delta_h^2(k),
    \label{GWspec}
\end{equation}
where $\Delta_h^2(k)$ is \cite{Odintsov:2021kup},
\begin{equation}\label{mainfunctionforgravityenergyspectrum}
    \Delta_h^2(k)=\Delta_h^{({\rm p})}(k)^{2}
    \left ( \frac{\Omega_m}{\Omega_\Lambda} \right )^2
    \left ( \frac{g_*(T_{\rm in})}{g_{*0}} \right )
    \left ( \frac{g_{*s0}}{g_{*s}(T_{\rm in})} \right )^{4/3} \nonumber  \left (\overline{ \frac{3j_1(k\tau_0)}{k\tau_0} } \right )^2
    T_1^2\left ( x_{\rm eq} \right )
    T_2^2\left ( x_R \right ),
\end{equation}
while the function $\Delta_h^{({\rm p})}(k)^{2}$ denotes the
inflationary tensor power spectrum, which is equal to
\cite{Odintsov:2021kup},
\begin{equation}\label{primordialtensorpowerspectrum}
\Delta_h^{({\rm
p})}(k)^{2}=\mathcal{A}_T(k_{ref})\left(\frac{k}{k_{ref}}
\right)^{n_{T}}\, .
\end{equation}
We shall evaluate the tensor power spectrum  at the CMB pivot
scale $k_{ref}=0.002$$\,$Mpc$^{-1}$ and the parameter $n_{T}$ is
the tensor spectral index, while $\mathcal{A}_T(k_{ref})$ denotes
the amplitude of the tensor perturbations,
\begin{equation}\label{amplitudeoftensorperturbations}
\mathcal{A}_T(k_{ref})=r\mathcal{P}_{\zeta}(k_{ref})\, .
\end{equation}
Also with $r$ we denote the tensor-to-scalar ratio and also
$\mathcal{P}_{\zeta}(k_{ref})$ is the amplitude of the primordial
scalar perturbations. Combining the above, we have,
\begin{equation}\label{primordialtensorspectrum}
\Delta_h^{({\rm
p})}(k)^{2}=r\mathcal{P}_{\zeta}(k_{ref})\left(\frac{k}{k_{ref}}
\right)^{n_{\mathcal{T}}}\, ,
\end{equation}
therefore the energy spectrum of the primordial gravitational
waves for an inflationary theory with tensor-to-scalar ratio $r$
and tensor spectral index $n_T$ is,
\begin{align}
\label{GWspecfR}
    &\Omega_{\rm gw}(f)=\frac{k^2}{12H_0^2}r\mathcal{P}_{\zeta}(k_{ref})\left(\frac{k}{k_{ref}}
\right)^{n_{\mathcal{T}}} \left ( \frac{\Omega_m}{\Omega_\Lambda}
\right )^2
    \left ( \frac{g_*(T_{\rm in})}{g_{*0}} \right )
    \left ( \frac{g_{*s0}}{g_{*s}(T_{\rm in})} \right )^{4/3} \nonumber  \left (\overline{ \frac{3j_1(k\tau_0)}{k\tau_0} } \right )^2
    T_1^2\left ( x_{\rm eq} \right )
    T_2^2\left ( x_R \right )\, ,
\end{align}
with $T_{\rm in}$ denoting the horizon reentry temperature
\cite{Odintsov:2021kup},
\begin{equation}
    T_{\rm in}\simeq 5.8\times 10^6~{\rm GeV}
    \left ( \frac{g_{*s}(T_{\rm in})}{106.75} \right )^{-1/6}
    \left ( \frac{k}{10^{14}~{\rm Mpc^{-1}}} \right )\, ,
\end{equation}
and in addition, the transfer function $T_1(x_{\rm eq})$ is
\cite{Odintsov:2021kup},
\begin{equation}
    T_1^2(x_{\rm eq})=
    \left [1+1.57x_{\rm eq} + 3.42x_{\rm eq}^2 \right ], \label{T1}
\end{equation}
with $x_{\rm eq}=k/k_{\rm eq}$ and $k_{\rm eq}\equiv a(t_{\rm
eq})H(t_{\rm eq}) = 7.1\times 10^{-2} \Omega_m h^2$ Mpc$^{-1}$,
and moreover, the transfer function $T_2(x_R)$ is
\cite{Odintsov:2021kup},
\begin{equation}\label{transfer2}
 T_2^2\left ( x_R \right )=\left(1-0.22x^{1.5}+0.65x^2
 \right)^{-1}\, ,
\end{equation}
where $x_R=\frac{k}{k_R}$.  Furthermore, the wavenumber when the
reheating temperature is achieved is,
\begin{equation}
    k_R\simeq 1.7\times 10^{13}~{\rm Mpc^{-1}}
    \left ( \frac{g_{*s}(T_R)}{106.75} \right )^{1/6}
    \left ( \frac{T_R}{10^6~{\rm GeV}} \right )\, ,  \label{k_R}
\end{equation}
with $T_R$ denoting the reheating temperature. Moreover,
$g_*(T_{\mathrm{in}}(k))$ is equal to,
\begin{align}\label{gstartin}
& g_*(T_{\mathrm{in}}(k))=g_{*0}\left(\frac{A+\tanh \left[-2.5
\log_{10}\left(\frac{k/2\pi}{2.5\times 10^{-12}\mathrm{Hz}}
\right) \right]}{A+1} \right) \left(\frac{B+\tanh \left[-2
\log_{10}\left(\frac{k/2\pi}{6\times 10^{-19}\mathrm{Hz}} \right)
\right]}{B+1} \right)\, ,
\end{align}
with the parameters $A$ and $B$ standing for,
\begin{equation}\label{alphacap}
A=\frac{-1-10.75/g_{*0}}{-1+10.75g_{*0}}\, ,
\end{equation}
\begin{equation}\label{betacap}
B=\frac{-1-g_{max}/10.75}{-1+g_{max}/10.75}\, ,
\end{equation}
where $g_{max}=106.75$ and $g_{*0}=3.36$. In addition,
$g_{*0}(T_{\mathrm{in}}(k))$ and can it can be calculated by
combining Eqs. (\ref{gstartin}), (\ref{alphacap}) and
(\ref{betacap}), by simply replacing $g_{*0}=3.36$ with
$g_{*s}=3.91$. At this point, we shall evaluate the effect of the
$z\sim 3400$ total EoS oscillations on the $h^2$-scaled energy
spectrum of the primordial gravitational waves. As we showed in
the previous section, the total EoS oscillates between the values
$w_1=0.2$ and $w_2=0.13$. The redshift $z=3400$ corresponds to
wavenumbers in the range $k\sim 10^{-2}-10^{-1}$Mpc$^{-1}$, thus
we shall consider the scenario for which the total EoS value
$w_1=0.2$ occurs at the wavenumber $k_1=9\times 10^{-2}$Mpc$^{-1}$
and the value $w_2=0.13$ occurs at the wavenumber $k_2=0.5\times
10^{-1}$Mpc$^{-1}$. Of course in reality this procedure of
oscillations will involve more than two wavenumbers, and we shall
consider several distinct scenarios here. If the total EoS is
deformed to a value $w$ at a wavenumber $k_s$, the energy spectrum
of the primordial gravitational waves is multiplied by the factor
$\sim \left(\frac{k}{k_{s}}\right)^{r_c}$, with
$r_c=-2\left(\frac{1-3 w}{1+3 w}\right)$
\cite{Gouttenoire:2021jhk}. Therefore, for an oscillation of the
total EoS in the range $w_1=0.2$-$w_2=0.13$ the $h^2$-scaled
energy spectrum of the primordial gravitational waves takes the
final form,
\begin{equation}\label{GWspecfRnewaxiondecay}
\Omega_{\rm gw}(f)=S_k(f)\times
\frac{k^2}{12H_0^2}r\mathcal{P}_{\zeta}(k_{ref})\left(\frac{k}{k_{ref}}
\right)^{n_{\mathcal{T}}} \left ( \frac{\Omega_m}{\Omega_\Lambda}
\right )^2
    \left ( \frac{g_*(T_{\rm in})}{g_{*0}} \right )
    \left ( \frac{g_{*s0}}{g_{*s}(T_{\rm in})} \right )^{4/3} \nonumber  \left (\overline{ \frac{3j_1(k\tau_0)}{k\tau_0} } \right )^2
    T_1^2\left ( x_{\rm eq} \right )
    T_2^2\left ( x_R \right )\, ,
\end{equation}
with $S_k(f)$,
\begin{equation}\label{multiplicationfactor1}
S_k(f)=\left(\frac{k}{k_{1}}\right)^{r_{s_1}}\times
\left(\frac{k}{k_{2}}\right)^{r_{s_2}}\, ,
\end{equation}
where $k_1=9\times 10^{-2}$Mpc$^{-1}$ and $k_2=0.5\times
10^{-1}$Mpc$^{-1}$ while $r_{s_1}=-2\left(\frac{1-3 w_1}{1+3
w_1}\right)$ and $r_{s_2}=-2\left(\frac{1-3 w_2}{1+3 w_2}\right)$.
Now we can explicitly confront the predictions of the effect of
the $F(R)$ gravity generated total EoS oscillations on the energy
spectrum of the primordial gravitational waves, and we shall
consider three reheating temperatures, a high reheating
temperature of the order $T_R=\mathcal{O}(10^{12})$GeV, an
intermediate one of the order $T_R=\mathcal{O}(10^{7})$GeV and a
low reheating temperature of the order $T_R=\mathcal{O}(5\times
10^{2})$GeV. As it proves, the reheating temperature does not
affect such low frequencies, so we choose only one reheating
temperature for our final analysis. Also we added some more
scenarios, the case of two oscillations, three and four
oscillations,  in which case the total EoS varies as follows, for
$k_1=9\times 10^{-2}$Mpc$^{-1}$, $w_1=0.13$, for $k_2=0.5\times
10^{-1}$Mpc$^{-1}$ we have $w_2=0.2$, for $k_3=0.5\times
10^{-2}$Mpc$^{-1}$ we have $w_3=0.13$ and for $k_4=0.1\times
10^{-2}$Mpc$^{-1}$ we have $w_4=0.2$. In which case the factor
$S_k(f)$ in Eq. (\ref{GWspecfRnewaxiondecay}) would be changed to,
\begin{equation}\label{multiplicationfactor1newshow}
S_k(f)=\left(\frac{k}{k_{1}}\right)^{r_{s_1}}\times
\left(\frac{k}{k_{2}}\right)^{r_{s_2}}\times
\left(\frac{k}{k_{3}}\right)^{r_{s_3}}\times
\left(\frac{k}{k_{4}}\right)^{r_{s_4}}\, ,
\end{equation}
where $r_{s_i}=-2\left(\frac{1-3 w_i}{1+3 w_i}\right)$. Now let us
proceed to the predictions of the theoretical framework we
developed for the energy spectrum of the primordial gravitational
waves. In Fig. \ref{FRMAIN} we present the energy spectrum of the
$F(R)$ gravity theory with $z\sim 3400$ total EoS, versus the
sensitivity curves of various current and future gravitational
waves experiments, for two, three and four oscillations of the
total EoS in the range $w=[0.13,0.20]$.
The two oscillations case corresponds to the red line, the three
oscillations case is depicted with the blue line, while the four
oscillations case is depicted with the magenta line. As it can be
seen, the energy spectrum of the primordial gravitational waves is
significantly enhanced for frequencies probed by the future
LiteBIRD space mission which will start operating in 2029. Now
this is particularly interesting, since if the spectrum of the
tensor perturbations is enhanced in the operation frequencies of
the LiteBIRD mission, this will mean that the CMB tensor modes
will be significantly enhanced for modes corresponding to these
frequencies and a detection of the B-mode spectrum could be
ensured. This feature characterizes all the three scenarios we
examined, with the most pronounced scenario being the four
oscillations case. Now let us discuss in brief this LiteBIRD
frequency enhancement due to the total EoS oscillations for
redshifts $z\sim 3400$. The fact that the energy spectrum of the
primordial gravitational waves is enhanced for frequencies probed
by the future LiteBIRD mission, this means plainly spoken that
such a signal can affect the CMB and might be detectable by
LiteBIRD but only under specific conditions. The frequencies
correspond to to extremely long-wavelength gravitational waves
that stretch over the largest observable scales in the Universe.
These waves directly affect the CMB by generating B-mode
polarization through their quadrupolar effect on the photons from
the last scattering surface. The frequencies probed by LiteBIRD,
correspond to CMB modes with angular scales on the sky around
multipole moments $\ell \sim 2-10$, which are large angular scale
modes. LiteBIRD is designed to measure exactly this range of
angular scales (and beyond), making it sensitive to such signals
if they are strong enough. The detection by LiteBIRD of such
gravitational waves depends on the amplitude of the gravity wave,
if the gravity wave signal produces a strong enough
tensor-to-scalar ratio then  LiteBIRD could detect it in terms of
a B-mode in the CMB. LiteBIRD aims to measure significantly small
tensor-to-scalar ratios thus if an enhancement occurs above the
operational frequencies, LiteBIRD will detect it. Apart from the
enhanced B-mode spectrum in the CMB, the enhanced gravitational
wave energy spectrum can in principle enhance the temperature
anisotropies at large angular scales, but the B-modes are the
clearest signature of the tensor perturbations. There are a
handful of theories which predict such a gravitational wave
pattern in the energy spectrum of the primordial gravitational
waves, and one of these theories is the present $F(R)$ gravity
framework. It is however challenging to discriminate these
theories between them.

\section{Concluding Remarks}

The underlying theory governing the evolution of the Universe must
have several inherent appealing features and characteristics. It
must be particle physics motivated, and must uniquely describe the
early and late-time acceleration eras in a unified way. Such a
theoretical framework is $F(R)$ gravity, which is known to provide
a unified and elegant description of inflation and the dark energy
era. The functional form of the $F(R)$ gravity can be chosen
purely on a phenomenological basis, but there exist a class of
exponential deformations of standard $R^2$ gravity which stems
from theoretical requirements related to the scalaron mass for de
Sitter spacetime perturbations \cite{mywork}. In this work we
studied in depth one such exponential deformation of $R^2$
inflation, and we addressed both the inflationary and dark energy
aspects of this model. As we showed, the exponential $R^2$
deformation yields a standard $R^2$ inflationary era with a
rescaled Einstein-Hilbert term. As we explicitly demonstrated, the
rescaling of the Einstein-Hilbert term does not affect at all the
$R^2$ inflationary dynamics, thus the inflationary era is
described by a standard $R^2$ inflationary era, which is viable
and well fitted in the Planck data. Moreover, the model at late
times provides a viable dark energy era which is compatible with
the Planck constraints on the cosmological parameters and emulates
the $\Lambda$CDM model regarding the behavior of several
statefinder quantities. Although the very late time behavior of
the $F(R)$ model mimics the $\Lambda$CDM model, near the epoch of
matter-radiation equality the $F(R)$ gravity model deviates from
the $\Lambda$CDM model behavior and strong total EoS oscillations
occur for redshifts $z\sim 3400$. This is a late-time era
oscillating period for the total EoS and it can affect the energy
spectrum of the primordial waves for very small frequencies. We
studied in detail the effect of the total EoS oscillations on the
energy spectrum of the primordial gravitational waves and we found
a significant enhancement of the energy spectrum for frequencies
that will be probed by the LiteBIRD mission. This enhancement can
affect the CMB and thus make possible the detection of the
$B$-mode in the CMB. The energy spectrum for the specific model we
studied has a characteristic pattern of primordial gravitational
waves, thus a detection of the $B$-mode spectrum by the LiteBIRD
and a simultaneous non-detection by the Simons observatory can
mean only one thing: The spectrum of the primordial gravitational
waves is enhanced near the matter-radiation equality, or that some
other mechanism near that era might have enhanced the energy
spectrum. We thus offered a new possibility for theories producing
gravitational waves. Note that in this framework, no higher
frequencies detections of primordial gravitational waves occur, so
this is indeed a unique pattern. Of course it is possible that
other mechanisms combined with the above scenario may yield
detectable gravitational waves by LISA and other missions, for
example phase transitions during the reheating era, but this is
something to be discussed in another work, in conjunction with the
present scenario.

Now in the present article we did not consider collective effects
of modified gravities containing higher order curvature
corrections in terms of parametric approaches, or effects like the
Sachs-Wolfe effects on the CMB, which is very much related to the
frequency range probed by the LiteBird, as for example in
\cite{Capozziello:2007vd}. In addition, several early phase
transitions can also contribute to the frequency range discussed
in this paper, like for example \cite{Capozziello:2018qjs}, these
topics must be considered in conjunction with the discussions made
in this article. However let us note that currently we are in the
process of understanding enhancement mechanisms of primordial
gravitational waves, and when the stochastic cosmological signal
is finally detected, we hopefully be at a position to judge which
mechanism generates the detected pattern of primordial
gravitational waves.

\section*{Acknowledgements}

This work was partially supported by the program Unidad de
Excelencia Maria de Maeztu CEX2020-001058-M, Spain (S.D.O). This
research is funded by the Committee of Science of the Ministry of
Education and Science of the Republic of Kazakhstan (Grant No.
AP26194585) (Vasilis K. Oikonomou).

\end{document}